# Leggett's plasma resonances and two-gap structures in the CVCs of $MgB_2$ break junctions – a direct evidence for a two-gap superconductivity in $MgB_2$

Ya.G. Ponomarev[1], S.A. Kuzmichev[1], M.G. Mikheev[1], M.V. Sudakova[1], S.N. Tchesnokov[1], N.Z. Timergaleev[1], A.V. Yarigin[1],

M.A. Hein[2], G. Müller[2], H. Piel[2],

B.M. Bulychev[3], K.P. Burdina[3], V.K. Gentchel[3], L.G. Sevastyanova[3],

S.I. Krasnosvobodtsev[4], A.V. Varlashkin[4]

[1] M.V. Lomonosov Moscow State University, Faculty of Physics, 119991 Moscow, Russia
[2] Bergische Universität Wuppertal, Fachbereich Physik, D-42097 Wuppertal, Germany
[3] M.V. Lomonosov Moscow State University, Faculty of Chemistry, 119991 Moscow, Russia
[4] P.N. Lebedev Physical Institute, RAS, 119991 Moscow, Russia

# PREFACE

## On the occasion of 20th anniversary of the first experimental observation of the Leggett collective plasma oscillation in $MgB_2$

Before proceeding to the electron copy of the original publication from December 2002 [1] that was never available on-line, I want to sketch a general problem of Leggett collective plasma oscillation and make several remarks both about the nature of phenomena that was firstly observed by Prof. Yaroslav Georgievich Ponomarev in the spectra of tunneling contacts and published in the Mendeleev University bulletin [1] and on the history of our very first publications.

Superconductivity in $MgB_2$ was discovered occasionally in the end of 2000 [2]. Despite the strong boron isotope effect (observed in [3]) clearly points to the classical phonon nature of the pairing mechanism in $MgB_2$, it becomes the first-ever-known two-gap superconductor (SC), which means that two types of the Cooper pairs having the distinct coupling energies ($2\Delta_1$ and $2\Delta_2$) are developed in the SC state below $T_c$. These condensates are not totally independent: they weakly interact through interband coupling in the momentum space. This is somehow similar to the proximity effect between two SC in real space, but in the former case they induced their intrinsic properties to each other in every point of crystal in the momentum space (*k*-space). As a result, the SC order parameter of one of the condensates ($\Delta_j$) being dependent both from his eigen properties (intraband coupling constant $\lambda_{jj}$), as well as from the strength of the crossband coupling $\lambda_{12}$ or $\lambda_{21}$ (see works by Moskalenko [4] and Suhl *et al.* (so-called SMW-model) [5] done independently back in 1959). Even been initially in the weak-coupling limit of BCS theory (e.g. $\lambda_{11}, \lambda_{22} < 0.25$ and having characteristic ratios $2\Delta^{\text{eigen}}(0) / k_B T_c = 3.53$), the variation of the coupling potential $V_{ij}$ in the intraband and crossband channels ($V_{\text{intra}} \neq V_{\text{inter}}$) leads to the $2\Delta_1(0) / k_B T_c > 3.53 > 2\Delta_2(0) / k_B T_c$ deviation (when $\Delta_1$ is the large SC gap) in case of an extension of the BCS theory for two bands (so-called two-band BCS).

Why this compound having the simple chemical formula $MgB_2$ demonstrates the variety of sophisticated physics? One of the reasons is in its layered crystal structure and the complexity of the Fermi surfaces (at least a couple of 2D-hole barrels which are nearly orthogonal with two 3D electron and hole constructions [6–8]). The latter is in the contrast with the conventional SC materials (so-called BCS superconductors) having more or less isotropic crystal structure, conductivity and 3D-electron Fermi surfaces. In occasion of the SC state, Cooper pairs with the same properties (at least coupling energy $2\Delta$) are developed at any and all conductive bands of the classical SC due to the crossband (interband) mixing of momenta. While this classical phenomenon being one of the important consequence of the Philip W. Anderson theorem, the magnesium diboride breaks its concept down in case of the 2D Fermi surfaces, crystal structure anisotropy and weak interband interaction. More detailed formulation one can find in [9].

Anthony J. Leggett (you can see several intriguing notes on his biography and "Reflections on the past, present and future of condensed matter physics" in [10]) in his pioneer work [11] predicted that collective oscillation of a superconducting plasma, which are caused by small fluctuations of the phase difference between two superconducting condensates, develop in two-gap (and at least two-band) superconductors. Two type of oscillations become possible: the in-phase "Goldstone" mode $\omega_G(k)$ that demonstrates acoustic-sound-like dispersion on the wave-vector $k$, and the out-of-phase "Leggett" mode shows exciton-like massless $\omega_L(k)$ dependence.

Charge carriers can flow from one band another band creating a crossband AC current in the $k$-space having some characteristic frequency $\omega_L(k)$. In his Nobel lecture [12] Leggett calls this type of the collective excitations as "a sort of internal Josephson effect", since they are intrinsic to the superconductor, and think that he was inspired by the "P.W. Anderson's elegant formulation of the theory of superfluidity in $^4$He in terms of conjugate number and phase variables". He also remembered that in 1966 his theory "sank more or less without trace, in part because by the time it was published it had already become clear that the experimental evidence for the existence of two-band superconductors in nature was dubious" [12].

The main result of Anthony Leggett's theory [11] is that the square of the oscillation frequency $\omega_L(k)$ of the exciton-like mode slightly depends on the wave-vector $k$ and is determined by two terms: the threshold frequency $\omega_0^2$ and the $u^2k^2$ product. For this reason in the $k \to 0$ limit, $\omega_L(0) = \omega_0$ and may be referred to as the massless term. It was shown that $\omega_0$ does not directly depend on the Coulomb interaction and could be obtained even for a system of neutral particles [11]. Finally,

$$\omega_0^2 = 4\Delta_1(0)\Delta_2(0)\frac{\lambda_{12}+\lambda_{21}}{\lambda_{11}\lambda_{22}-\lambda_{12}\lambda_{21}} \qquad (1)$$

and valid in the limits of (a) $T \to 0$ strictly, (b) for small wave-vectors $k$, (c) in the low-energy limit (let say BCS weak-coupling constants $\lambda_{ij} < 0.25$, while theoretical estimations [6] give $\lambda_{11} \approx 1$, $\lambda_{22} \approx 0.3$), (d) resulting Leggett plasma frequency $\omega_0$ must correspond to the in-gap energies $\omega_0 \leq 2\Delta_2(0) < 2\Delta_1(0)$, so as not to be strongly damped by the quasiparticle continuum.

This result was re-derived by Sharapov *et al.* [13] specially for $MgB_2$. The numerical estimation made in [13] demonstrate the energy range $2\Delta_2(0) < \omega_0 < 2\Delta_1(0)$ for the $\omega_0$, contrary do the clause (d) limitations. Also note that a value of the exciton-type mode estimated for $MgB_2$ in [13] is $\omega_0 \approx \Delta_1(0) + \Delta_2(0)$ that rises additional difficulties for experimentalists to distinguish between the Leggett and the threshold cross-gap energy excitations, if it would be observed in an experiment.

The latter raised a question on the possibility of the experimental observation of the Leggett mode, since it should be seriously damped. The limitation of $\omega_0$ by the smallest SC gap value $2\Delta_2$ (so-called softening of the Leggett mode or anticrossing with the gap edge) become a point of the theoretical discussions, for example, see Eq. 2 and Fig. 3 in the work by Karakozov *et al.* [14]; see also Figs. 3,4 from [15]. In the latter paper Klimin *et al.* argue that "The low frequency expansion thus becomes inapplicable when the Leggett mode frequency approaches the range close to the pair-breaking continuum edge." Slightly above this sentence it has been written: "[it] does not capture the interplay of the Leggett collective mode with the pair-breaking continuum edge and hence crosses the value $\omega = 2\Delta_2$ without any feature." [15]. The same problem has been addressed in nice theoretical exercises of Arimitsu [16] (see Fig.1).

What is the most important result of [11] for the experiment is the linear dependence of the $\omega_0^2$ on the $\Delta_1(0)\Delta_2(0)$ product (in the low-energy limit, and $T \rightarrow 0$) given by Eq.(1). The same direct scaling was obtained in the number of theoretical studies [14,17–19]. The direct proportionality $\omega_0^2 \sim \Delta_1(0)\Delta_2(0)$ can be checked by the gaps $\Delta_j$ variation with doping in $(Mg,Al)B_2$ and $Mg(B,C)_2$ systems. From the other side, this verification can help to distinguish between $\Delta_1\Delta_2$ Leggett's and threshold cross-gap $(\Delta_1+\Delta_2)^2$ dependencies of the $\omega_0^2$.

The story of the experimental discovery of Leggett collective mode started early in 2001, while the author of these notes began to work on his PhD thesis in Lomonosov MSU under the supervision of Prof. Ya.G. Ponomarev. His tunneling effects laboratory was developed in 1986 (as a consequence of the discovery of high-temperature SC cuprates). Ya.G. Ponomarev had extended classical "break-junction" tunneling setup of Moreland and Ekin [20] to be used with layered single crystals and realized the mechanically controlled planar "break-junction" (MCP-BJ) technique to produce *S-c-S* (*S* – bulk superconductor, *c* – constriction) contacts in *ab*-plane [21,22] and bulk natural arrays *S-c-S-c-...-c-S*. MCP-BJ technique should be used to study namely layered superconductors materials, see also our brief review [23]. Sadly, Prof. Ponomarev passed away in December 2015 after a severe and prolonged illness.

Already in the January of 2001 we get samples of a newly discovered superconductor $MgB_2$ made in the group of Prof. B.M. Bulychev from the Chemical faculty of Lomonosov Moscow State University (MSU). This occasion resulted in a change of the postgraduate work plan and its aim for the author. Our laboratory of tunneling effects has gone deep into the study of magnesium diboride electron properties. No one could guess at the time that we were dealing with the first two-gap (or two component) SC! Tunneling features of rather large amplitude at low bias region (caused by the small $\Delta_\pi$-gap) annoyingly entered the dI(V)/dV-curves, and for several months we tried to get rid of them, so that they did not "spoil" our spectra.

In the beginning of the summer we have got new series of $MgB_2$ samples from the Chemical faculty of MSU made in high-pressure chamber by Bulychev, Burdina and Sevastyanova with the different level of a structural disorder, as well as samples with the special made admixture of Mg-oxide (up to ~10%) produced by the magnesium vaporization method by Krasnosvobodtsev and Varlashkin (Lebedev Physical Institute, Russian Academy of Sciences). The latter samples had unprecedented properties: a width of the resistive transition to the SC state was as small as $\Delta T \approx 0.2$ K (i.e. ~0.5% of $T_c$), and, in addition, they demonstrated unusually large $\sigma$-gap values and $T_c$'s that reached 39–41 K (compare it with the standard maximum $T_c \approx 38.5$ K)! But that is another interesting topic not related to the Leggett mode.

During summer of 2002, in the dynamic conductance spectra of planar tunnel contacts (in both regimes, low-transparent and low-capacity Josephson SIS-contacts, and high-transparent semiballistic or diffusive Andreev SNS-contacts with incoherent transport and weak inelastic scattering [24–26]) based on $MgB_2$, as well as $MgB_2$ + MgO samples, Ya.G. Ponomarev discovered some reproducible additional fine structure corresponding to (a) the resonant excitation of some boson mode by the AC Josephson current in the range of energies that corresponds to small SC gap $2\Delta_\pi$ for SIS-contacts (with the current deficiency), and (b) to the excess loss of energy (due to the multiple boson emission) by the so-called Andreev carriers (electrons involved in multiple Andreev reflections (MAR) [24–26]) in the bands with large gap $2\Delta_\sigma$ for the high-transparent SnS-contacts (with the excess current and so-called “foot” structure at low bias [24–26]). It is interesting to compare this phenomenon observed in $MgB_2$ with the single or multiple spin-exciton resonant emission during MAR in SnS-contacts based on Fe-based SC of 1111 family (on the issue of the boson-mode) observed by us [27–29], see also the scheme of the emission process in Fig. 3 of [29].

Yaroslav Georgievich Ponomarev told us to search for the special looking fine structure at dI(V)/dV-spectra for observing its reproducibility in our tunneling break-junctions (note, only *bulk* properties or effects are reproducible in randomly shaped break-junctions!) and for discovering its temperature dependence. He was the first man, who compares the characteristic energy of both additional peculiarities (fine structures) in Josephson and Andreev transport regimes (SIS- *vs*. SNS-contacts data, as well as data for the corresponding SISIS and SNSNS arrays), ties it together and understands the same Leggett mode nature of the both effects.

During 2002 the reproducibility of these effects was observed a dozen times and more or less verified for $MgB_2$ having the largest $T_c \approx$ 35–40 K. For sure, the nature of the boson resonances found by Prof. Ponomarev required clarification, and we had to check it out. One of the most probable situations (both for transport and optical measurements) could be *indirect* tunneling of quasiparticles from the top of a valence band #1 to the bottom of a conductive band #2 (in this case they definitely change their band due to the inelastic process during tunneling). This should give something like threshold crossband excitation energy $(\Delta_\sigma+\Delta_\pi)$ ~ 8–11 meV value for $MgB_2$ with a critical $T_c$ ~ 34–38 K.

Contrary to these expectations, we reproducibly observed the half of these values (4–5 meV) on the one hand, and on the other hand, the realization of the indirect (cross-gap) tunneling in our planar ScS contacts would definitely produce the (large amplitude) fundamental gap structure at $eV = \Delta_\sigma + \Delta_\pi$, but not the additional fine-structure, as we have observed. Since that, we concluded in the Leggett mode nature of the resonances observed, and Prof. Ponomarev decided to make the experimental results public.

The first results were published in the very end of 2002 in the "Bulletin of the Mendeleev University of Chemistry and Technology", see Ya.G. Ponomarev *et al.* [1]. The conference proceedings makes possible to quickly publish the results, which played a positive role, since the next paper on this issue was sent to an editorial office of a high-impact physical journal during the spring of 2003 [30], and its publication was continuously postponed by referees. Finally, in September 2003, the manuscript was readdressed to the editorial office of "Solid State Communications", and was immediately accepted as a "hot topic publication". It become available online in October 2003, but physically appears just in the $2^{nd}$ issue in January 2004 [31], thus, formally, this led to the loss of two years, since the experimental discovery of the phenomenon predicted back in 1966. Note that the "arXiv:cond-mat" version [30] has color figures vs. black-and-white graphs in Solid State Communications [31].

It was very gratifying that Anthony J. Leggett referred to this work [31] in his Nobel lecture [12]. Aside from the pleasure to experimentally discover some new phenomena, we remained a little skeptical and curious, could this experimental resonant energy be driven something else than Leggett collective excitation?

In result, we checked, how does this resonant energy $\omega_L(0)$ vary with aluminum doping of $Mg_{1-x}Al_xB_2$ and, correspondingly, its $T_c$? Since we can measure both bulk SC gaps (directly at $T \to 0$), thus, we checked the linear relationship between (momentum independent part of) the Leggett collective excitation energy squared and a product of the experimental values of the σ-gap and the π-gap: $\omega_0^2 \sim \Delta_\sigma(0)\Delta_\pi(0)$, according to the Leggett's equation (1) in the wide range of critical temperatures $10\ K < T_c < 40.5\ K$.

The result of this important experimental verification showed $\omega_0^2 \approx \Delta_\sigma \Delta_\pi$ (at $T = 4.2\ K << T_c$) and was firstly published in Fig. 6 of Ya.G. Ponomarev, *et al.*, "Leggett's mode in $Mg_{1-x}Al_xB_2$" [32], and several years later in Fig. 2 of [14], as well as in paragraph 4.2 of [33]. Unfortunately, so far, we can not find any experimental work done by optical methods, in which this energy $\omega_0$ would be measured by the Raman response on doped $MgB_2$ together with SC gaps $\Delta_j$ and $T_c$ variation. This issue is still waiting to be checked by optical spectroscopy.

Subjecting self-criticism to the work of Ponomarev's laboratory, in which I was fortunate enough to participate, I need to mention such a shortcoming as the lack of discussion on the Andreev bound states (ABS) influence on our SNS-contact spectra. The development of ABSs in “long” SNS junctions [25, 34–38] has generally the same physical origin as the quantum size effect in (normal) metallic films as a result of the superposition of incident and reflected electron waves [39], as well as the Tomasch size effect in SIS/N tunneling structures [40], in which low energy carriers ($|\varepsilon| \leq \Delta$) participate in the (single) Andreev reflection at the S/N-interface of the structure, reversing, among other parameters, the sign of their charge. This leads to interference of the incident electron and reflected hole waves in ‘S’ at a distance of the order of mean-free-path, resulting in a series of peculiarities formation in the local electron density of states (DOS).

In case of SNS contact low energy carriers are involved in MAR process, being the ground state of this tunneling system, yet at $eV \to 0$ bias voltage. They produce the single, several or even a comb of the nearly equidistant ABS *inside* the SC gap (i.e. in the “forbidden” range of energies), depending on a ratio of a metal layer width $d$ to the SC coherence length $\xi_0$ (see [41] for some details). Electrons and holes are prohibited from entering the energy range within the SC gap inside the SC, but not into the normal metal layer ‘N’. As the ratio $d/\xi_0$ is increase, more and more ABS appear *inside* the gap region, producing new maxima of local DOS, until a bunch of Andreev levels merge into a zone and the influence of the proximity of bulk ‘S’ to thin ‘N’ will stop.

According to the theories of MAR effect [24–26] the most energetic is the first Andreev minima in dI(V)/dV-spectra of SNS contact (the so-called fundamental harmonic) that is biased at $eV_1 = 2\Delta$. Consequently, any of the *in-gap* features, including ABS, have to appear at $eV^* < 2\Delta$ [25]. Contrary to this, we have repeatedly observed extra features at energies *large* than $2\Delta_\sigma$, for example, see minima marked as “m=1”, “m=2”, “m=3” in Fig. 6 of [1,30], or the same in Fig. 5 of [31] (note that label “$n_L$=1” points to $2\Delta_\sigma$ fundamental minima). These additional minima arise from the phenomena of the multiple boson emission by the Andreev carriers. The position (bias) of the $m$ = 1,2,3… fine structure defined by (or from the point of view of the experimentalist, define) the trivial expression: $eV_{1,\mathrm{m}} = 2\Delta + m\omega_0$ that demonstrates definitely *overgap* energies and leaving no chance to be originated from the comb of ABS.

We thank P.I. Arseev and N.K. Fedorov for several short but fruitful discussions during 2010–2013, as well as A.V. Galaktionov for advices on the problem of ABS.

S.A. Kuzmichev,

December 2022

*The electron copy of paper* [1] *is presented below.*

**Mendeleev Chemical Society of Russia**
**Russian Physical Society**
**Mendeleev University of Chemical Technology of Russia**

# Bulletin of the V. Tarasov Center of Chemotronics of Glass
# No. 2

## *PREFACE*

*August 21 this year is the 100th anniversary of Vasilii Vasil'evich Tarasov, an outstanding scientist and brilliant pedagogue who headed the Chair of Physics at the Moscow Mendeleev University of Chemical Technology. Numerous generations of scientists and specialists who studied at the Mendeleev University recall the lectures by V.V.Tarasov as most memorable impressions of their student years that determined their path into science.*

*N.F.Mott, V.V.Tarasov, R.L.Müller and B.T.Kolomiets are undoubtedly the pioneers who played the key role in the chemotronics of glass. This is a systematic concept that considers behavior and properties of glass on the basis of elementary processes and mechanisms that cannot be divided into purely physical or chemical. From the point of view of chemotronics of glass, these processes include electrical conductivity, brittle fracture, viscous flow, switching effect, phenomena in lightguides. Chemotronics of glass is also a basis for understanding glass as a self-organizing system with memory that can be used for information storage.*

*A scientific symposium "Problems of chemotronics of glass" dedicated to Tarasov's 100th anniversary took place at the Mendeleev University of Chemical Technology on October 10—11, 2002. Both well-known and young scientists from Moscow, St. Petersburg and other Russian scientific centers took part in the symposium. This second issue of the Bulletin of the V.V.Tarasov center of the chemotronics of glass includes the papers by the symposium participants. The Bulletin also contains some papers by well-known Russian scientists on the basic problems of science presented by the invitation of the Organizing Committee. The Organizing Committee thanks all who took part in this memorable event of Russian science.*

*Chairman of the Organizing Committee of the Symposium*
*President of the Tarasov Center of Chemotronics of Glass*
*Member of the Russian Academy of Science*

*P.D.Sarkisov*

начальную магнитную проницаемость, уменьшает коэрцитивную силу, снижая при этом термостабильность. С позиций предлагаемой гипотезы становится понятным наблюдаемое уменьшение $\mu_H$ и рост $K_П$ при увеличении константы текстуры в образцах. Аналогичная картина наблюдалась и на ферритах 1500 НМЗ, МГП. Легко объяснить и низкую эффективность ТМО магнитно-текстурованных ферритов. Существование анизотропных деформаций, препятствуя процессу направленного упорядочения, не позволяет достичь желаемого эффекта. При этом, как было показано выше, в магнитно-текстурованных образцах, как правило, невелика концентрация ионов $Fe^{2+}$ , что также затрудняет процесс создания направленного порядка. Поэтому для повышения эффективности термомагнитной обработки необходимо избавиться от магнитной текстуры в образцах или свести ее к минимуму. Таким образом, предлагаемое объяснение возникновения магнитной текстуры в марганцевых ферритах полностью согласуется с результатами проведенного эксперимента.

Выдвинутая гипотеза возникновения магнитной текстуры в поликристаллических Mn-Fe-Zn ферритах согласуется с результатами рентгеноспектральных исследований пол и кристаллических ферритов этой системы, показавшими присутствие в них янтеллеровских ионов $Mn^{3+}$. Выполненная оценка возможного вклада ионов $Mn^{3+}$ в магнитную анизотропию подтвердила разумность предложенной модели.

# Leggett's plasma resonances and two-gap structures in the CVCs of $MgB_2$ break junctions – a direct evidence for a two-gap superconductivity in $MgB_2$

**Ya.G. Ponomarev, S.A. Kuzmitchev, M.G. Mikheev, M.V. Sudakova, S.N. Tchesnokov, N.Z. Timergaleev, A.V. Yarigin.**
M.V. Lomonosov Moscow State University, Faculty of Physics.
**M.A. Hein, G. Müller, H. Piel.**
Bergische Universität Wuppertal, Fachbereich Physik, D-42097 Wuppertal, Germany.
**B.M. Bulychev, K.P. Burdina, V.K. Gentchel, L.G. Sevastyanova,**
M.V. Lomonosov Moscow State University, Faculty of Chemistry.
**S.I. Krasnosvobodtsev, A.V. Varlashkin.**
P.N. Lebedev Physics Institute, RAS.

## 1. Introduction

Theoretical and experimental investigations of the nature of high-temperature superconductivity are far from completion [1-4]. However, due to studies of high-temperature superconductors (HTSC), that employed the most modern experimental methods, an enormous body of data has been gathered and theoretical models for describing the unique properties of HTSCs have been built. Note that even today there is no agreement in the choice of the pairing mechanism [5-8].

Doubts in the universal nature of the magnon pairing mechanism [8] in HTSC appeared after the discovery of a new superconductor, magnesium diboride $MgB_2$ with a critical temperature $T_c$ = 39 K [9]. The pairing mechanism in $MgB_2$ is of phonon nature, which is proved by the discovery in this compound of an isotope effect [10]. The B isotope substitution ( $^{11}B \rightarrow {}^{10}B$) shifts $T_c$ by about 1K, while the Mg isotope effect is ten times smaller.

Theoretical analysis of the band structure of $MgB_2$ and related compounds showed that the conduction along the boron planes is close to two-dimensional ($\sigma$-bands) [11]. The presence of a van Hove singularity in the 2D band may strongly affect the value of $T_c$ if one shifts the Fermi level to the peak in the quasi-particle density of states through doping [12].

According to a popular version [11, 13], magnesium diboride displays a two-gap superconductivity, and at T= 4.2 K the gap $\Delta_L \cong 7$ meV corresponds to the 2D charge carriers in the $\sigma$-bands, while the gap $\Delta_S \cong 2$ meV corresponds to 3D carriers in the $\pi$-bands. Calculations have shown that both gaps close simultaneously at the critical temperature $T_c \cong 40$ K, with the temperature dependence of both gaps close to the standard BCS dependence. The theoretical quasi-particle density of states has two distinctive gap singularities, which must result in two independent subharmonic gap structures, corresponding to $\Delta_L$ and $\Delta_S$, appearing in the current-voltage characteristics of Andreev point contacts of SnS type. Accordingly two-gap structures are expected in the CVCs of tunneling NIS and SIS junctions.

In 1966 Leggett had predicted for superconductors with two bands of charge carriers a collective oscillation mode corresponding to small fluctuations of the relative phases of the two superconducting condensates [14, 15]. An expression for the energy of the Leggett's plasma mode for $MgB_2$ has been derived by Sharapov, Gusynin and Beck [16]:

$$E_0^2 = 4\Delta_L\Delta_S[(\lambda_{12}+\lambda_{21})/(\lambda_{11}\lambda_{22} - \lambda_{12}\lambda_{21})] \quad \text{.....(1)},$$

where $\lambda_{ij}$ - dimensionless interband and intraband coupling constants. The estimated values of the Leggett's plasmon energy $E_0$ lie in the range from 6.5 meV to 8.9 meV [16].

In principle this plasma mode should be observable with electromagnetic radiation, but in practice the effect on the infrared optical properties of a two-gap superconductor is too small to be observable [17 ]. At the same time a Josephson junction on the basis of a two-gap superconductor can be used to detect a collective plasma mode originally proposed by Leggett [18]. A resonance enhancement of the DC current through a Josephson junction at bias voltage $V_l$ is expected when the Josephson frequency $\omega_J$ or its harmonics $(l \cdot \omega_J)$ match the energy of the Leggett's mode $E_0$[18 ]:

$$E_0 = 2elV_l \; (l \text{ – is an integer number}) \quad \text{.....(2).}$$

If the experiment is done at finite current this will show up as Fiske steps.

In case of Andreev point contacts of the SnS type the resonant emission of Leggett's plasmons with the energy $E_0$ would cause the appearance of several sets of subharmonic gap structures at bias voltages:

$$V_{n,m} = (2\Delta_L + mE_0)/en, \quad \text{.....(3),}$$

where n is an integer number and m is a number of emitted Leggett's plasmons.

In the present investigation the current-voltage characteristics (CVCs) of break junctions in polycrystalline $MgB_2$ samples have been studied in the temperature range

$4.2K \leq T \leq T_c$. An inelastic Cooper pair tunneling accompanied by the emission of Leggett's plasmons (equation 2) has been found for the first time in $MgB_2$ Josephson contacts. A fine structure in the CVCs of $MgB_2$ Andreev point contacts caused by a resonant emission of Legget's plasmons (equation 3) has been observed for the first time. An energy of the Leggett's plasmon has been estimated : $E_0 \cong 4$ meV. Distinct two-gap structures have been found in the CVCs of $MgB_2$ tunneling contacts and $MgB_2$ Andreev point contacts with $7.5 \text{ mev} \leq \Delta_L \leq 11$ meV and $1.5 \text{ meV} \leq \Delta_S \leq 2.5$ meV. An intrinsic tunneling effect (ITE) and intrinsic multiple Andreev reflections effect (IMARE) have been observed due to the layered structure of $MgB_2$. The obtained experimental results give a direct evidence for a two-gap superconductivity in $MgB_2$.

## 2. Experimental results obtained in $MgB_2$ studies using the Josephson, tunneling and Andreev spectroscopy methods

When applied to $MgB_2$, the methods of Josephson, tunneling and Andreev (point-contact) spectroscopies demonstrated their efficiency and made it possible to extract useful information about the physical properties of this material in the superconducting state. Below we briefly discuss some recent experimental results of tunneling and point- contact measurements involving $MgB_2$ samples.

In the present investigation a comparative study of superconducting properties of three sets of $MgB_2$ polycrystalline samples has been performed. The first two sets of $MgB_2$ samples (BG series and BBS series) were prepared by Bulychev, Gentchel (Faculty of chemistry, MSU) and Bulychev, Burdina and Sevastyanova (Faculty of chemistry, MSU) respectively. The third set of $MgB_2$ samples (KV series) has been prepared by Krasnosvobodtsev and Varlashkin (Physics Institute, RAS). Different technique of preparation has been used. For BG series the resistive transition started at $T_{c,start} = 39$ K and finished at $T_{c,\,fin} = 29$ K. For BBS series the resistive transition started at $T_{c,\,start} = 40.5$ K and finished at $T_{c,\,fin} = 40.2$ K. The same is true for KV series.

The following experimental methods were employed in our investigations:

(1) Andreev spectroscopy (multiple Andreev reflections in $MgB_2$ break junctions of the SnS type);

(2) tunneling spectroscopy (single Josephson $MgB_2$ SIS junctions);

(3) intrinsic tunneling spectroscopy - ITS (intrinsic Josephson effect in microsteps on the cryogenic cleavage surface of $MgB_2$);

(4) intrinsic multiple Andreev reflections effect (IMARE) in microsteps on the cryogenic cleavage surface of $MgB_2$ samples.

All these methods of investigation of the superconducting properties of $MgB_2$ involve using a break junction technique. The transition from one measurement mode to another was done by mechanically tuning the break junction at the liquid-helium temperature.

### 2.1. Intrinsic tunneling (Josephson) effect and intrinsic multiple Andreev reflections effect

It should be noted that there are substantial specific difficulties in fabricating **single** tunneling and Andreev junctions in $MgB_2$ with current in **c**-direction (**j** || **c**). These difficulties are related to the layered nature of the $MgB_2$ crystal structure: in **c**-direction the banks of a tunneling junction are themselves a natural stack of resistively shunted Josephson junctions (**σ**-bands). The role of shunt in k-space play **π**-bands .

In the present investigation the intrinsic tunneling (Josephson) effect and the intrinsic multiple Andreev reflections effect were observed on the natural ultrathin steps, which are always present on a cryogenic cleavage surface of $MgB_2$ samples (Fig. 1 and Fig. 2). In Fig. 1 the dI/dV-characteristics of two stacks of five SIS contacts normalized to a single contact (BG series, T = 4.2 K, curve 1 and curve 2) are compared to normalized dI/dV-characteristic of a stack of two SIS contacts (BG series, T = 4.2K, curve 3) and to dI/dV-characteristic of a single SIS contact (BBS series, T = 4.2 K, curve 4). In Fig. 2 normalized dI/dV-characteristics of stacks of SnS Andreev contacts (BG series, T = *4.2 K, curve 1 – five contacts, curve 2 – two contacts) are compared to dI/dV-characteristic of a single SnS Andreev contact (BG series, T = 4.2 K, curve 3).*

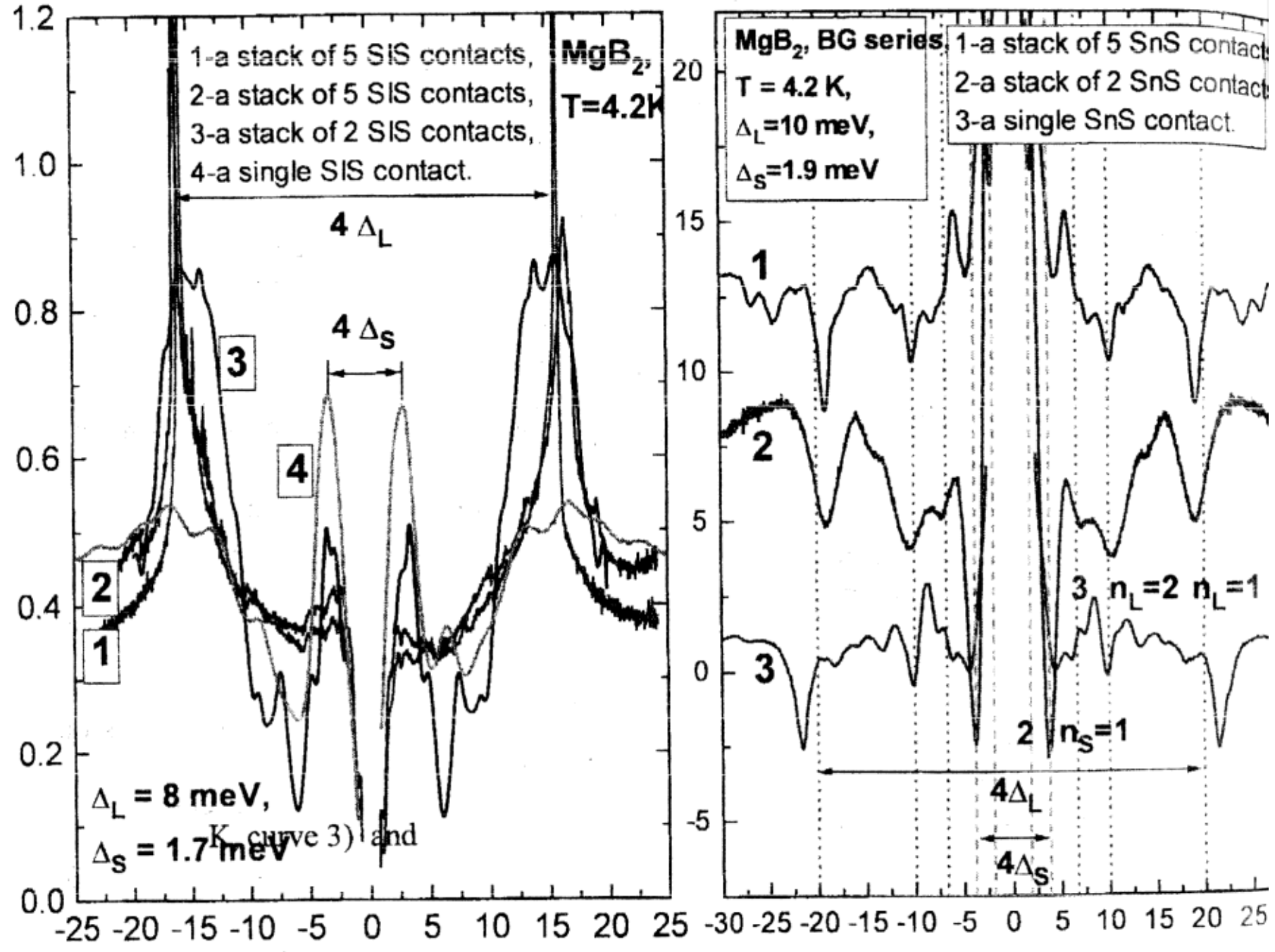

*Fig. 1. A two-gap structure in normalized CVCs of SIS $MgB_2$ contacts (T = 4.2 K).*

*Fig. 2. Two sets of SGS with $\Delta_L$ and $\Delta_S$ in normalized CVCs of Andreev contacts.*

In Fig. 1 a clear two-gap structure with $\Delta_L = (8 \pm 0.5)$ meV and $\Delta_S = 1.7$ meV is present in all curves. In Fig. 2 two sets of subharmonic gap structures (SGS) are detectable with $\Delta_L = (10 \pm 1)$ meV and $\Delta_S = 1.9$ meV.

*Until recently, the intrinsic Josephson effect was observed only in cuprate HTSC.*

*2.2. Tunneling and Andreev spectroscopies. Calculating superconducting gaps $\Delta_L$ and $\Delta_S$*

*In the present investigation the gap structure in the dI/dV-characteristic of a junction in the tunneling regime (the peak value of the differential conductance for a gap voltage $V_g = 2\Delta/e$) has been compared to the subharmonic gap structure (SGS) in the dI/dV-characteristic of the same junction in the point-contact (Andreev) mode*

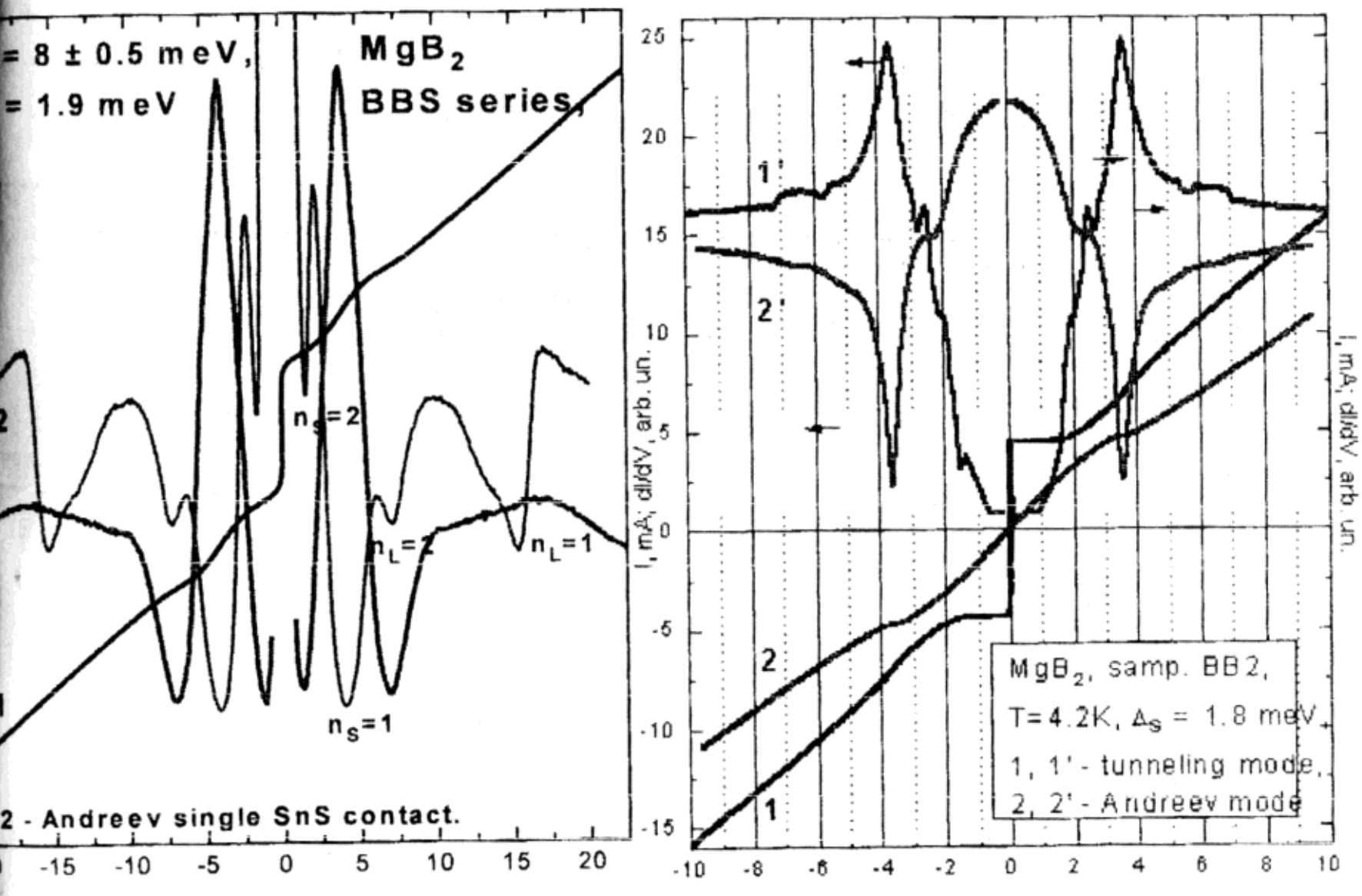

*Fig. 3. Two-gap structure in the CVCs of $MB_2$ break junctions in the tunneling regime (1, 1') and Andreev regime (2).*

*Fig.4. $\Delta_S$ - structure in the CVCs of $MB_2$ break junctions in the tunneling regime (1,1') and Andreev regime (2,2').*

*(a series of dips in the differential conductance at bias voltages $V_n = 2\Delta/en$, where n is an integer) (Fig. 3 and Fig. 4). Earlier Muller et al. [19] did the same type of research with niobium break junctions. The value of the superconducting gap $\Delta$ was assumed reliable only if the values of $\Delta$ obtained by the above two methods were equal.*

*In the present investigation the current-voltage characteristics of moze than 150 Andreev point contacts and tunneling contacts were studied in the temperature interval from 4.2 K to $T_c$. As a first step, the histograms representing the dependence of the number of junctions on the value of the superconducting gap at 4.2 K had been plotted for the BG and KV series of $MgB_2$ samples. Both histograms exhibit pronounced peaks at least at three values of the gap: $\Delta_1$, $\Delta_2$, and $\Delta_3$. In the case of the BG series, $\Delta_1 = (2.0 \pm 0.5)$ meV, $\Delta_2 = (8.0 \pm 0.5)$ meV, and $\Delta_3 = (16.0 \pm 0.5)$ meV; in the case of the KV series, $\Delta_1 = (2.0 \pm 0.5)$ meV, $\Delta_2 = (10.5 \pm 0.5)$ meV, and $\Delta_3 = (21.0 \pm 0.5)$ meV. According to our interpretation, the gap $\Delta_2$ corresponds to a large two-dimensional gap $\Delta_L$ ($\sigma$ - bands) [11 - 13] and the gap $\Delta_3$ to $2\Delta_L$, which is possible in the case of stacks of two Andreev or tunneling junctions (the consequence of the layered structure of $MaB_2$). The gap $\Delta_1$ corresponds to a small three-dimensional gap $\Delta_S$ ($\pi$ – bands) [11 - 13].*

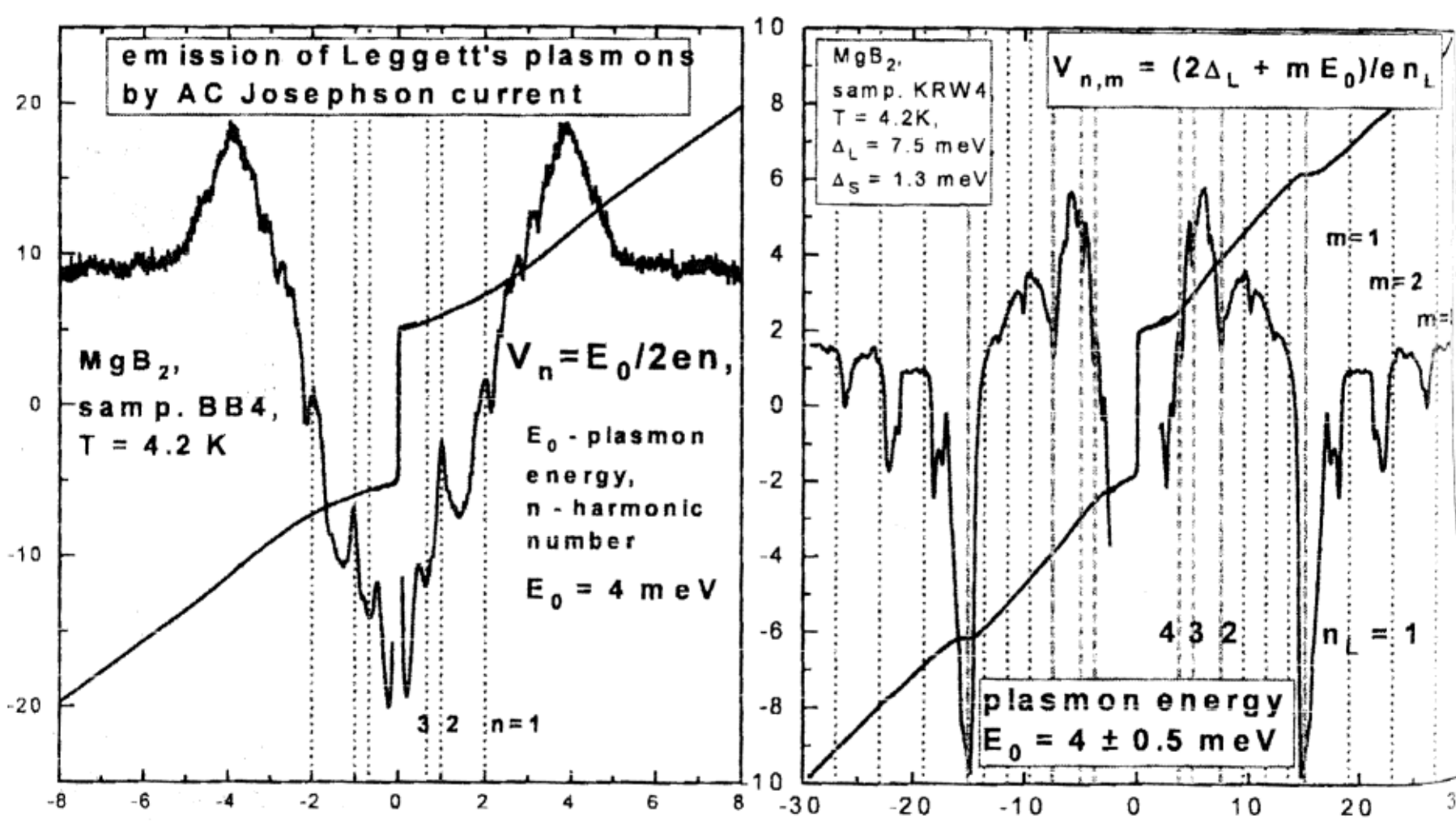

*Fig. 5. Structure in the CVC of a SIS $MgB_2$ junction caused by generation of Legget's plasmons ($T$=4.2K, $E_0$=4meV)*

*Fig.6 Subharmonic gap structure modified by emission of plasmons.*

We have found that the temperature dependence of the large gap $\Delta_L$ is described by the BCS model with the ratio $2\Delta_L/kT_c$, amounting to 6.0 ± 0.5 for the BG series and 6.5 ± 0,5 for the KV series. These values are close to $2\Delta_L/kT_c$ for superconducting cuprates. The temperature dependence of the small gap $\Delta_S$ differs substantially from the BCS model, and the ratio $2\Delta_S/kT_c$ has no fixed value. The above facts suggest that the interband scattering (**σ-π** transitions) greatly affects the size of the small gap $\Delta_S$.

*2.3. Josephson and Andreev spectroscopies. Resonant emission of Leggett's plasmons in $MgB_2$ break junctions – a direct evidence for a two-gap superconductivity in $MgB_2$.*

In the present investigation a **reproducible** fine structure in dI/dV-characteristics of Josephson $MgB_2$ junctions has been found for the first time (Fig. 5). This structure resembles qualitatively a structure in dI/dV-characteristics of HTSC Josephson junctions, which is caused by inelastic tunneling of Cooper pairs accompanied by generation of nonequilibrium optical phonons in the frequency range up to 20 THz [20, 21]. If we assume that the structure in Fig. 5 is caused by a resonant emission of some excitations by AC Josephson current, the energy of these excitations should not exceed $E_0$ = 4 meV. Such a low self energy has nothing to do with the optical phonons in $MgB_2$. **At the same time its value is close to the calculated energy of the Leggett's plasma mode for $MgB_2$ [16] (see equation 1).**

A careful inspection of the fine structure in dI/dV-characteristic (Fig. 5) shows that a related structure in the CVC has a form of sharp current peaks, predicted for $MgB_2$ Josephson junctions which resonantly emit Leggett's plasmons [18 ]. **The fine structure is well reproducible which means that its parameters are governed by intrinsic properties of $MgB_2$ samples. This is true for Leggett's plasma mode (see equation 1).** At last the fine structure is detectable only at temperatures $T < T_c$, which means that it is closely connected with superconductivity in $MgB_2$. The position of singularities correlates well with the equation (2) (see Fig. 5). **Thus the most probable origin of the fine structure in Fig. 5 is the resonant emission of Leggett's plasmons with $E_0$ = 4 meV by the AC Josephson current.**

Additional evidence for this version comes from Fig. 6. A complex form of a subharmonic gap structure in Fig. 6 can be explained by a resonant emission of Leggett's plasmons ($E_0$ = 4 meV) by quasiparticles, undergoing multiple Andreev retroreflections at SN-interfaces of Andreev point contact (see equation (3) and Fig. 6).

**Acknowledgements**

We would like to thank P.I. Arseev, V.F. Gantmakher, Yu.M. Kagan, E.G. Maksimov and L.M. Fisher for their extremely useful discussions of the results of the present research. The work was made possible by partial financial support from the Scientific Council of the Russian ANFKS state-sponsored R&D program (the 'Delta' Project) and the Russian Foundation for Basic Research (Grant No. 02-02-17915).

## ПОВЕРХНОСТНЫЕ И ОБЪЁМНЫЕ ТЕКСТУРЫ СТИЛВЕЛЛИТА В СТЕКЛАХ СИСТЕМЫ $La_2O_3$-$B_2O_3$-$GeO_2$.

**В.Н.Сигаев, А.М.Даценко, С.Ю.Стефанович[1], А.О.Пожогин, В.И.Фертиков, Д.А.Захаркин[2], В.В.Сахаров[2]**

РХТУ им. Д.И.Менделеева
[1]Институт Физической Химии им. Л.Я.Карпова
[2] ВСЕРОССИЙСКИЙ НАУЧНО-ИССЛЕДОВАТЕЛЬСКИЙ ИНСТИТУТ ХИМИЧЕСКОЙ ТЕХНОЛОГИИ.

**Аннотация.**
В стеклах системы $La_2O_3$-$B_2O_3$-$GeO_2$ составов вблизи стехиометрии стилвеллита $LaBGeO_5$ при определенных условиях могут быть сформированы как поверхностные, так и объёмные текстуры, пронизывающие весь объём стекла и инициирующие в закристаллизованных стеклах нелинейно-оптические, сегнетоэлектрические и родственные сегнетоэлектричеству свойства. В данной работе с помощью сканирующей электронной микроскопии иллюстрируется процесс поверхностной и объемной ориентированной кристаллизации стилвеллитных стекол для различных условий синтеза. Как поверхностные прозрачные (нелинейно-оптические), так и объёмные (сегнето-, пироэлектрические) текстуры представляют собой совокупность ориентированных перпендикулярно поверхности образца игольчатых кристаллов стилвеллита $LaBGeO_5$.

**Введение.**
Нелинейно-оптические среды на основе стекла привлекают все большее внимание исследователей. Электрическая поляризация стекол [1-2], наноструктурирование стекол нецентросимметричными фазами [3-6] позволяют

Научное издание

Вестник
Центра хемотроники стекла им. В.В. Тарасова
№ 2

Selectable and updated **REFERENCES** of the original work

Electronic layout of color **FIGURES** of the original work

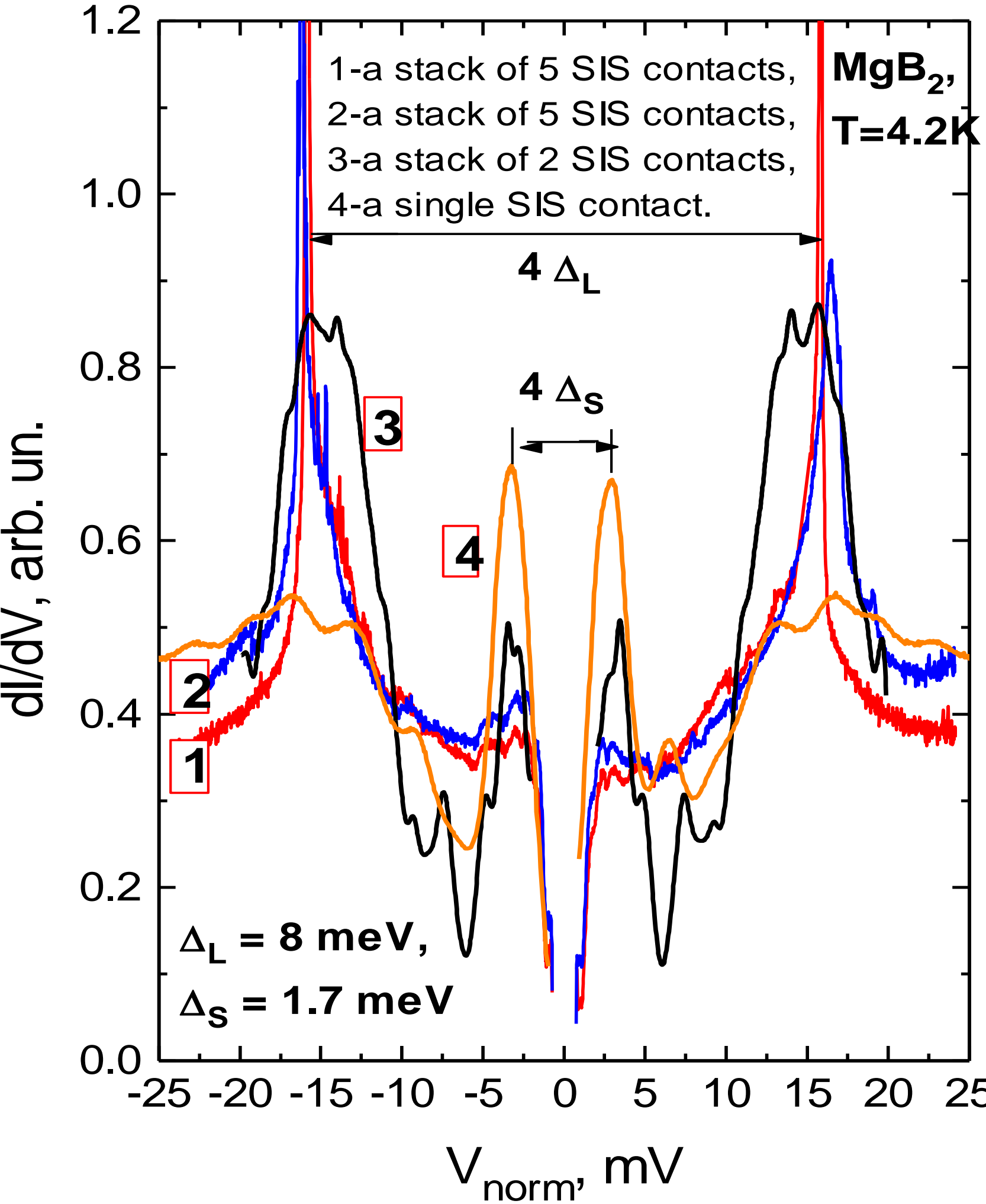

**Fig. 1**. A two-gap structure in normalized CVCs of SIS contacts based on $MgB_2$ (T = 4.2 K).

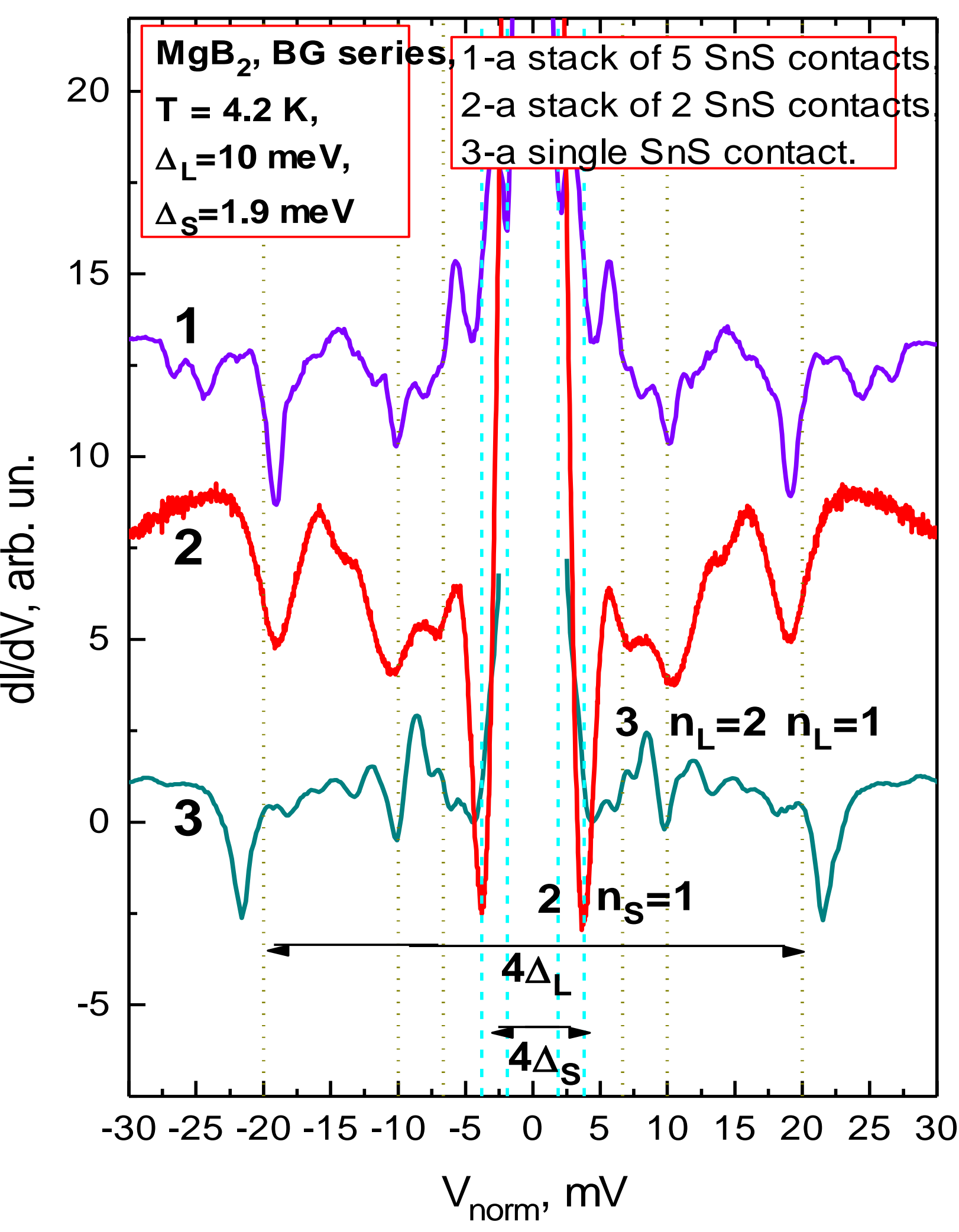

**Fig. 2**. Two sets of SGS with $\Delta_L$ and $\Delta_S$ in normalized CVCs of Andreev contacts.

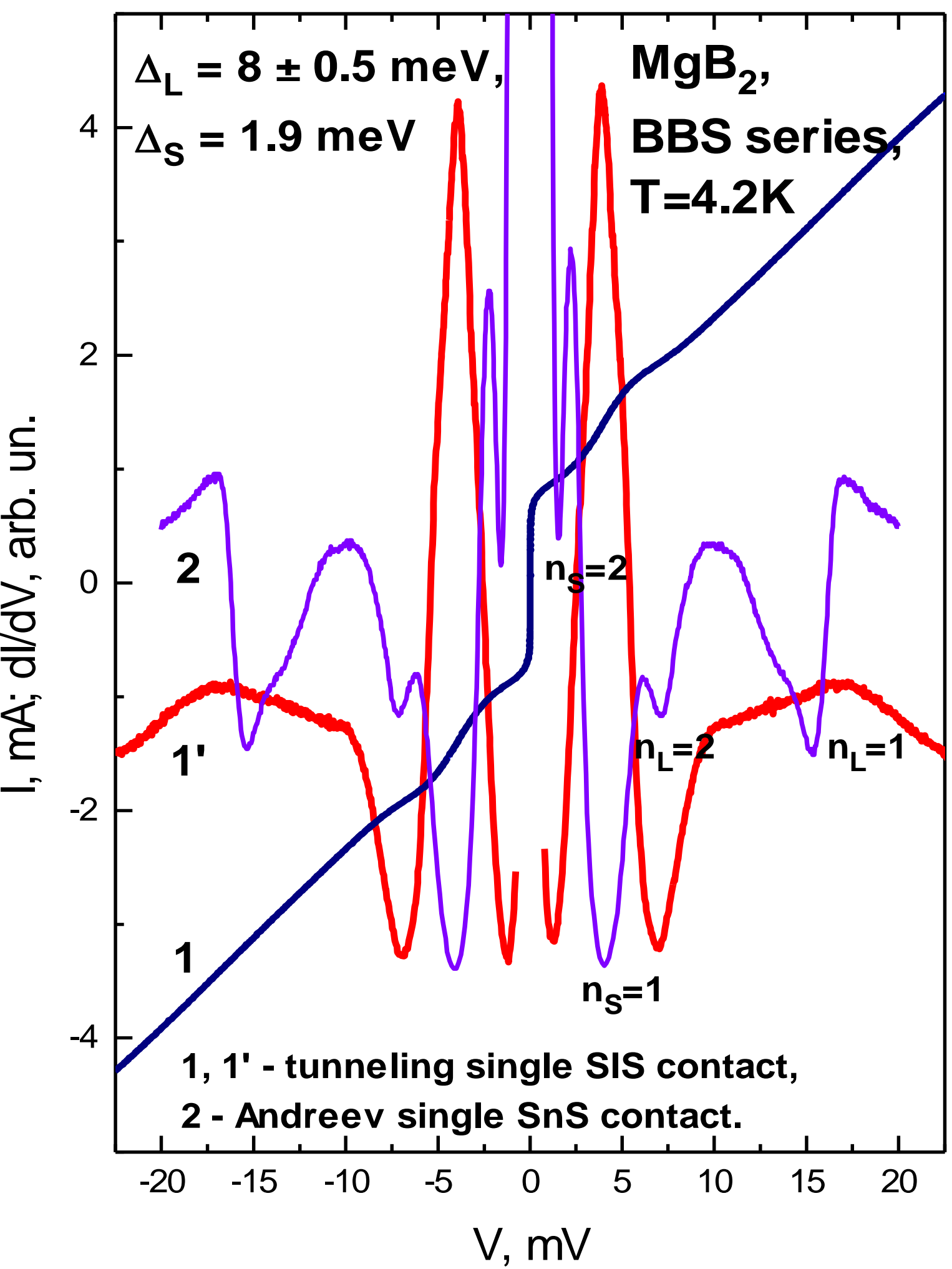

**Fig. 3**. Two-gap structure in the CVCs of $MB_2$ break junctions in the tunneling regime (1, 1') and Andreev regime (2).

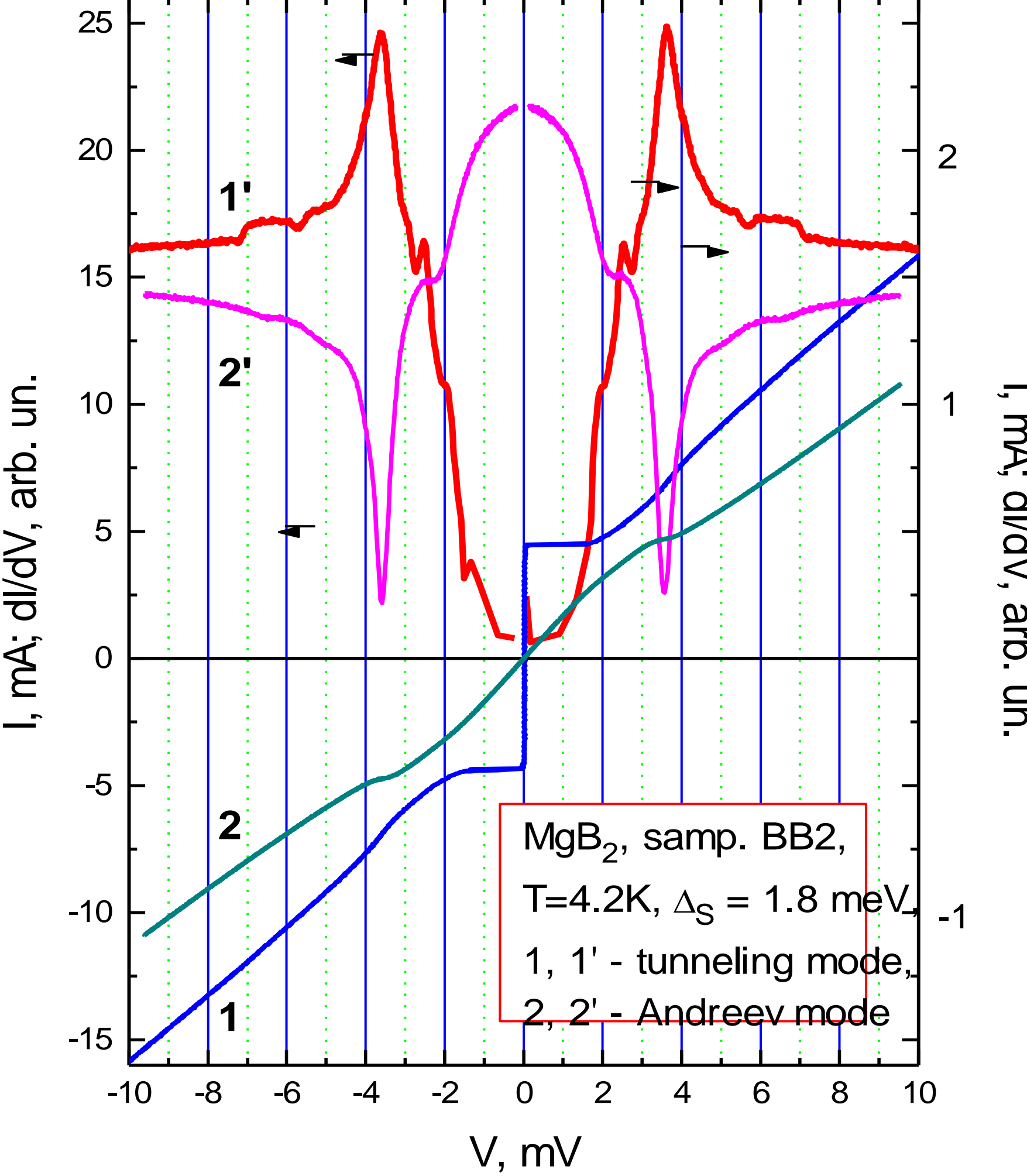

**Fig.4**. $\Delta_S$-structure in the CVCs of $MB_2$ break junctions in the tunneling regime (1,1') and Andreev regime (2,2').

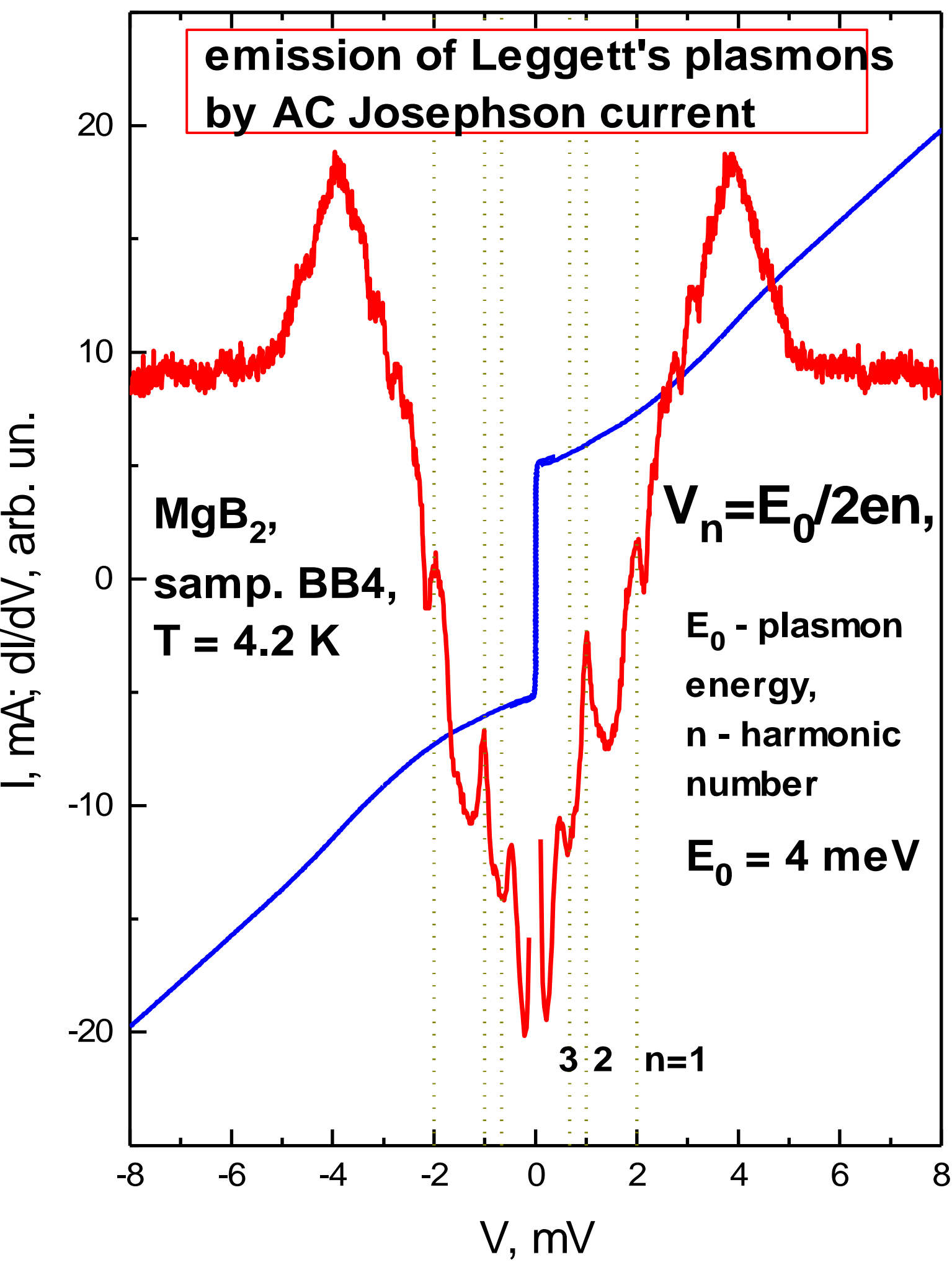

**Fig. 5**. Structure in the CVC of a SIS $MgB_2$ junction caused by generation of Leggett's plasmons (T = 4.2 K, $E_0$ = 4 meV).

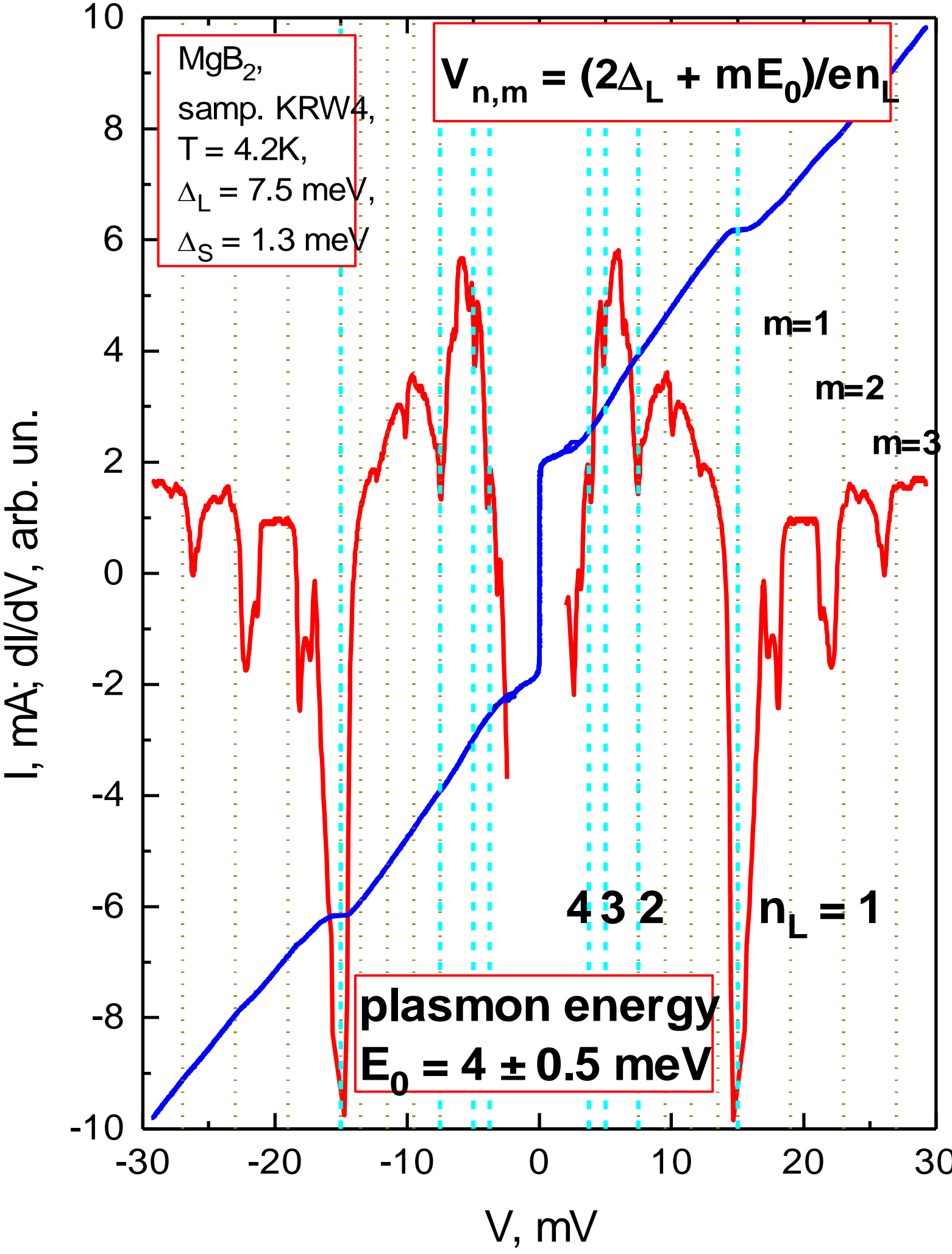

**Fig. 6**. Subharmonic gap structure modified by emission of plasmons.